\documentclass[
  aps,prl,
  superscriptaddress,
  twocolumn,
  showpacs,showkeys,
  amsmath,floatfix
]{revtex4-1}
\usepackage{graphicx}% Include figure files
\usepackage{dcolumn}% Align table columns on decimal point
\usepackage{bm}% bold math
\usepackage[
  dvipdfmx,colorlinks=true,linkcolor=blue,
  bookmarks=true,bookmarksnumbered=true
]{hyperref}% add hypertext capabilities
\usepackage{txfonts}
\usepackage{comment}
\begin{document}
%\preprint{}

\title{Direct determination of the neutron skin thicknesses
  in $^{40,48}$Ca \\
  from proton elastic scattering at $E_p = 295$ MeV}


\author{J.~Zenihiro}
\email[]{juzo@ribf.riken.jp}
%\thanks{}
\affiliation{RIKEN Nishina Center,
  Wako, Saitama 351-0198, Japan}
%%%%%
\author{H.~Sakaguchi}
\affiliation{Research Center for Nuclear Physics,
  Osaka University, Ibaraki, Osaka 567-0047, Japan}
\author{S.~Terashima}
\affiliation{School of Physics and Nuclear Energy Engineering,
  Beihang University, Beijing 100191, China}
%\author{Y.~Matsuda}
%\affiliation{Cyclotron and Radioisotope Center, Tohoku University,
%  Sendai, Miyagi, 980-8578, Japan}
\author{T.~Uesaka}
\affiliation{RIKEN Nishina Center,
  Wako, Saitama 351-0198, Japan}
%%%%%
\author{G.~Hagen}
\affiliation{Physics Division, Oak Ridge National Laboratory,
  Oak Ridge, Tennessee 37831, USA}
\affiliation{Department of Physics and Astronomy, University of Tennessee,
  Knoxville, Tennessee 37996, USA}
\author{M.~Itoh}
\affiliation{Cyclotron and Radioisotope Center, Tohoku University,
  Sendai, Miyagi 980-8578, Japan}
\author{T.~Murakami}
\affiliation{Department of Physics, Kyoto University,
  Kyoto 606-8502, Japan}
\author{Y.~Nakatsugawa}
\affiliation{Institute of High Energy Physics,
  Chinese Academy of Sciences, Beijing 100049, China}
\author{T.~Ohnishi}
\affiliation{RIKEN Nishina Center,
  Wako, Saitama 351-0198, Japan}
\author{H.~Sagawa}
\affiliation{RIKEN Nishina Center,
  Wako, Saitama 351-0198, Japan}
\affiliation{Center for Mathematics and Physics, University of Aizu,
  Fukushima 965-0001, Japan}
\author{H.~Takeda}
\affiliation{RIKEN Nishina Center,
  Wako, Saitama 351-0198, Japan}
\author{M.~Uchida}
\affiliation{Department of Physics, Tokyo Institute of Technology,
  Meguro, Tokyo 152-8551, Japan}
%\author{Y.~Yasuda}
%\affiliation{Research Center for Nuclear Physics,
%  Osaka University, Ibaraki, Osaka 567-0047, Japan}
\author{H.P.~Yoshida}
\affiliation{Research Center for Nuclear Physics,
  Osaka University, Ibaraki, Osaka 567-0047, Japan}
\author{S.~Yoshida}
\affiliation{Science Research Center, Hosei University,
  2-17-1 Fujimi, Chiyoda, Tokyo 102-8160, Japan}
\author{M.~Yosoi}
\affiliation{Research Center for Nuclear Physics,
  Osaka University, Ibaraki, Osaka 567-0047, Japan}
%\date{\today}

\begin{abstract}
  The neutron density distributions and neutron skin thicknesses
  in $^{40,48}$Ca are determined from the angular distributions
  of the cross sections and analyzing powers of polarized proton
  elastic scattering at $E_p = 295$ MeV.
  Based on the framework of the relativistic impulse approximation
  with the density-dependent effective $NN$ interaction,
  the experimental data is successfully analyzed, providing
  precise information of neutron and proton density profiles of
  $^{40,48}$Ca with small uncertainties.
  The extracted neutron and proton density distributions give neutron skin
  thicknesses in $^{40,48}$Ca for
  $-0.010^{+0.022}_{-0.024}$ fm and $0.168^{+0.025}_{-0.028}$ fm,
  respectively.
  The results of the density profiles and the neutron skin thickness
  in $^{48}$Ca are directly compared with the {\it ab initio} coupled-cluster
  calculations with interactions derived from chiral effective field theory,
  as well as relativistic and non-relativistic energy density functional theories.
\end{abstract}

%\pacs{
%  21.10.Gv,
%  21.65.Ef,
%  24.10.Jv,
%  25.40.Cm
%\keywords{}
\maketitle

%%%%%%%%%%%%%%%%%%%%%%%%%%%%%%%%%%%%%%%%%%%%%%%%%%%%%%%%%%%
%\paragraph*{Introduction}
%
Protons and neutrons in a nucleus tend to be distributed so that
%both densities overlap at their maximum and their sum
their density sum
does not exceed the saturation density $\rho_{\rm sat}\sim 0.17$~fm$^{-3}$.
The proton and neutron distributions are almost identical;
they have a plateau of density $\sim 0.5\rho_{\rm sat}$
at the center in symmetric nuclei where
the neutron number and the proton number are almost the same ($N \sim Z$).
On the other hand, excess neutrons in neutron-rich nuclei
are pushed to the nuclear surface, forming 
a region where only neutrons exist.
This region is called a ``neutron skin''.

Theoretical studies indicate that the thickness of the neutron skin
$\Delta r_{np}$, which is defined as the difference of the neutron and proton
root-mean-square (rms) radii ($\Delta r_{np} \equiv r_n - r_p$), embodies
the stability of pure neutron matter.
The quantity that characterizes the stability of neutron matter is called
``symmetry energy''.
The neutron matter equation of state (EOS) is a sum of well-known EOS
of the symmetric nuclear matter and the symmetry energy.
EOS governs not only the formation of nuclei but also
astrophysical phenomena like neutron stars and
super nova explosions. Consequently, the neutron matter EOS
has been intensively studied in both nuclear physics and astrophysics
\cite{Oertel2017, Radice2017, Roca-Maza2018, Fattoyev2018, Annala2018}.
The experimentally-measured $\Delta r_{np}$ can help elucidate
the nature of high-density neutron matter
occurring at neutron stars and binary neutron star mergers in the universe
\cite{Abbott2017}.
%Recently, the first merging event was observed with a gravitational wave
%\cite{Abbott2017}.

In previous studies, the doubly magic $^{208}$Pb has been used as a
benchmarking nucleus because the double magicity
removes the effects from a complicated nuclear structure.
This enables reliable comparisons between the experimental results
and theoretical predictions.
%The interrelationship between
%$\Delta r_{np}$ of $^{208}$Pb and the parameters in the symmetry energy
%has been theoretically investigated
%Calculations based on energy density functional (EDF) theories
%predict that
$\Delta r_{np}$ in $^{208}$Pb is theoretically predicted to have
a strong correlation with the coefficient $L$ of
the first density-derivative of the symmetry energy
\cite{Typel2001, Furnstahl2002, Chen2005a}.
Many facilities have made experimental efforts
to determine $\Delta r_{np}$ in $^{208}$Pb
by measuring proton elastic scattering \cite{Starodubsky1994b, Zenihiro2010},
coherent pion-photoproduction \cite{Tarbert2014a},
antiprotonic atom X-ray \cite{Kos2007},
and electric dipole polarizability \cite{Tamii2011a}.
%Energy density functional (EDF) theories
%predict that the electric dipole polarizability $\alpha_{D}$ 
%is correlated with the skin thickness, and the data of $\alpha_{D}$
%are used to evaluate the neutron skin thickness.
Their results are in the range of 0.15--0.21 fm with the error of
approximately $\pm$0.03 fm.
The PREX experiment using
parity-violating (PV) electron scattering resulted in
$\Delta r_{np} = 0.33^{+0.16}_{-0.18}$ fm \cite{Abrahamyan2012},
which is consistent with other results within its large statistical error.
The results have been compared with theoretical predictions
based on the relativistic and non-relativistic
energy density functional (EDF) theories
\cite{Roca-Maza2011, Tsang2012, Oertel2017, Roca-Maza2018}.

Recently,
%the neutron skin thickness
$\Delta r_{np}$ in $^{48}$Ca has been investigated
both theoretically and experimentally.
One merit of $^{48}$Ca is that the nucleus is within
the range of the state-of-the-art {\it ab initio} calculations 
\cite{Furnstahl2013, Soma2014, Ekstrom2015, Hergert2016, Ekstrom2018}.
An interesting physics case is the direct assessment of the three nucleon force
(3NF) effects. The 3NF plays an important role in
high-density nuclear matter \cite{Akmal1998}.
{\it Ab initio} coupled-cluster (CC) calculations
based on the chiral effective field theory (EFT) interactions
including the three-nucleon force, have been successfully performed
for $^{48}$Ca \cite{Ekstrom2015, Hagen2016a, Ekstrom2018}.
$\Delta r_{np}$ in $^{48}$Ca should exhibit
a new aspect of the neutron matter EOS, which cannot be seen
in $^{208}$Pb because an {\it ab initio} calculations cannot be
applied at present to $^{208}$Pb.
In this Letter, we present the results of the direct determination
of the neutron density distribution and
$\Delta r_{np}$ in $^{40,48}$Ca
by proton elastic scattering at 295 MeV.
The extracted neutron density distribution
and $\Delta r_{np}$ in $^{48}$Ca are compared with recent
{\it ab initio} CC calculations \cite{Hagen2015, Ekstrom2018} and
predictions of relativistic and non-relativistic EDF theories.

Several experimental studies on
$\Delta r_{np}$ in $^{48}$Ca have been performed and planned
\cite{Birkhan2017, Horowitz2014}.
The electric dipole polarizability $\alpha_D$ of $^{48}$Ca has been
precisely determined
by combining proton inelastic scattering data
with photoabsorption data \cite{Birkhan2017}.
$\Delta r_{np}$ in $^{48}$Ca has been deduced by comparing the data
with several EDF and {\it ab initio} theories.
%It should be noted that the relationship between $\Delta r_{np}$
%and $\alpha_D$ in medium-mass nuclei may be subject to
%larger model dependences than in the case of $^{208}$Pb.
%(e.g. in Ref \cite{}).
However, a direct determination of $\Delta r_{np}$ in $^{48}$Ca independent
from a specific nuclear structure model
is very important to examine the theory.
One possibility is a CREX experiment using PV electron scattering
which is a clean probe compared with hadronic probes
\cite{Horowitz2014}.
The electro-weak interaction mediated by a photon and the $Z^0$ boson
is used to probe neutrons inside a nucleus.
The experiment is quite difficult because the precision of ppb level is
required for the measurement of the PV asymmetry.
%In addition, for model-independent analysis of the density distribution,
%several data-sets of different momentum transfers are necessary.

Proton elastic scattering is another approach to
directly extract the neutron density distributions.
Compared to PV electron scattering, the advantages
%of the proton elastic scattering
are a high sensitivity to
the neutron density and a highly attainable statistics due
to a large cross section.
The cost for the high efficiency is more complicated reaction mechanism
than that in electron scattering.
An accurate determination of the neutron density distribution requires
an established reaction model and a reliable calibration using nuclei
with a well-known density distribution.

Herein we use proton elastic scattering at $\sim$300 MeV
and analysis in the framework of
the relativistic impulse approximation (RIA). 
The weakest nucleon-nucleon ($NN$) interaction appears around
300 MeV and the nucleus is the most transparent to an incident proton.
This allows the impulse approximation to be used for
the reaction analysis, providing the opportunity to 
determine the density distribution at the nuclear interior.

\begin{figure}[t]
  \begin{center}
  \includegraphics[scale=0.37]{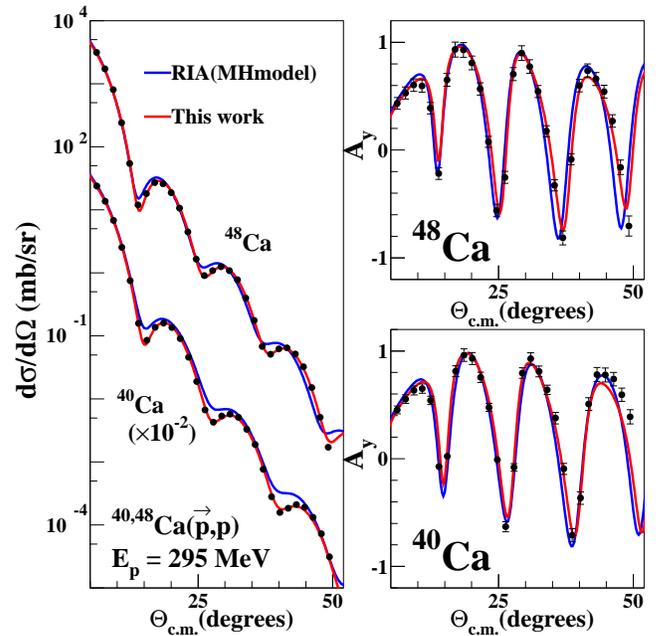}%
  \caption{
    Differential cross sections and analyzing powers for
    polarized proton elastic
    scattering from $^{40,48}$Ca at 295 MeV.
    Blue and red lines are from the MH model and
    the result of the best-fit in this analysis,
    respectively.
    \label{data48ca}}
  \end{center}
\end{figure}

%%%%%%%%%%%%%%%%%%%%%%%%%%%%%%%%%%%%%%%%%%%%%%%%%%%%%%%%%%%
%\paragraph*{Experiment}
The angular distributions of the cross sections and analyzing powers for
polarized proton elastic scattering from $^{40,48}$Ca were measured
with a high-resolution magnetic spectrometer, ``Grand Raiden''
\cite{Fujiwara1999}
at Research Center for Nuclear Physics, Osaka University,
over an angular range of $7^{\circ}$--$48^{\circ}$.
Polarized protons accelerated to 295 MeV were scattered by
a natural Ca foil (abundance of $^{40}$Ca 96.9\%)
and an enriched $^{48}$Ca foil (enrichment 97.7\%)
with thicknesses of 3.50 and 1.06 $\rm mg/cm^2$, respectively.
Beam polarization was kept above 70\% during the measurements.
Beam intensity was adjusted in the range of 1--400 nA,
depending on the scattering angle.
The momenta of scattered protons, which were analyzed by the Grand Raiden
spectrometer, were determined by the focal plane detectors.
The typical energy resolution was about 100 keV
in the full width at half maximum.
As shown in Fig. \ref{data48ca},
the measured data-sets of the differential cross sections and
analyzing powers of $^{40,48}$Ca($\vec{p},p$) (black dots)
cover
%were obtained over the wide angular range, corresponding to
the wide momentum transfer range of 0.5--3.5 fm$^{-1}$.

%%%%%%%%%%%%%%%%%%%%%%%%%%%%%%%%%%%%%%%%%%%%%%%%%%%%%%%%%%%
%\paragraph*{Extraction of the neutron density distributions
%  in $^{40,48}${\rm Ca}}
The neutron density distributions are determined by the method
employed in Refs \cite{Zenihiro2010, Sakaguchi2017}.
The method is based on the RIA model developed by Murdock and Horowitz
(MH model) \cite{MurdockDavidPandHorowitz1987}, which is among
the most successful models for proton elastic scattering
in the intermediate energy region ($\geq$ 200 MeV)
\cite{Sakaguchi2017}.
The present analysis modified the relativistic Love-Franey
(RLF) $NN$ interaction
employed in the original MH model \cite{Horowitz1985}
to take the effects of the nuclear medium into account,
e.g. Pauli blocking, multi-step processes,
and vacuum polarization effects.
%This modification is necessary to improve the description of
%proton scattering and to accurately extract
%the neutron density distributions.
The modification was introduced in a form of density-dependent
coupling constants and masses of mesons used in the RLF interaction
\cite{Sakaguchi1998}.
%In Ref. \cite{Sakaguchi1998}, Sakaguchi, {\it et al.} introduced
%the modification parameters
%in a form of density-dependent coupling constants and masses of
%mesons used in the RLF interaction.
%, which are expressed as
%\begin{eqnarray}
%  g_j^2,\ \bar{g}_j^2 &\rightarrow& 
%  \frac{g_j^2}{1+a_j\rho (r)/\rho _0},\
%  \frac{\bar{g}_j^2}{1+\bar{a}_j\rho (r)/\rho _0} \label{eq:MEg}\\ 
%  m_j,\ \bar{m}_j &\rightarrow& 
%  m_j\left[1+b_j\frac{\rho (r)}{\rho _0}\right],\
%  \bar{m}_j\left[1+\bar{b}_j\frac{\rho (r)}{\rho _0}\right] \label{eq:MEm}\\
%  && j=\sigma, \omega, \nonumber
%\end{eqnarray}
%where $g_j$, $\bar{g}_j$, $m_j$, and $\bar{m}_j$ are 
%the coupling constants and the masses of $\sigma$ and $\omega$ mesons
%for the real and imaginary amplitudes, respectively \cite{Sakaguchi1998}.
%$\rho(r)$ is the isoscalar density of the nucleus.
%The reference density $\rho_0$ is taken as 0.1934~fm$^{-3}$.
%In the parameterization, the masses and coupling constants of the exchanged mesons
%at $\rho(r)=0$ are the same as those in free space.
%The parameters $a_j, \bar{a}_j, b_j,$ and $\bar{b}_j$ were determined
As reported in Ref. \cite{Zenihiro2010},
the density-depndent parameters in the modification were well determined 
so that the medium-modified RIA calculation reproduces
the proton elastic scattering data
on $^{58}$Ni at the same incident energy as the present work.
%The reason that $^{58}$Ni is chosen as a reference is that
%the proton density is derived from
%the well-known nuclear charge distribution determined
%by electron scattering \cite{DeVries1987c}, and 
%the predicted neutron radius of $^{58}$Ni is
%almost the same as the proton radius
%in many mean field theories
%(e.g. in Refs. \cite{Hofmann2001, VRETENAR2005a, Sarriguren2007}).
%The neutron density distribution is assumed to have the same shape
%as the proton density distribution.
Previous studies \cite{Terashima2008, Zenihiro2010} have demonstrated
that the analysis method successfully works
in different regions of nuclei like tin and lead.
The parameters determined for $^{58}$Ni work well for $A=$ 116--208.
Thus, these parameters should be reasonable in the case of the calcium
isotopes with masses closer to $A=58$.

To single out the precise neutron density distribution $\rho_n(r)$,
the point-proton density distribution $\rho_p(r)$ is necessary.
The nuclear charge distribution $\rho_{ch}(r)$ was precisely determined from
the electron scattering data \cite{DeVries1987c}.
Thus, $\rho_p(r)$ can be derived by 
unfolding $\rho_{ch}(r)$ with the intrinsic proton and neutron
electromagnetic distributions.
Using the Fourier transforms of the distributions in the momentum space $q$ 
%($F_i(q)\equiv{\mathcal F}\{\rho_i(r)\}$),
%($F_i(q)\equiv{\mathcal F}\{\rho_i(r)\}=4\pi\int\rho_i(r)j_0(qr)r^2dr$),
%($F_i(q)\equiv{\mathcal F}\{\rho_i(r)\}=4\pi\int_0^{\infty}\rho_i(r)
%\tfrac{\sin(qr)}{(qr)}r^2dr$),
($F_i(q)\equiv{\mathcal F}\{\rho_i(r)\}=\int\rho_i(r)
\exp(i\pmb{q} \cdot \pmb{r})d\pmb{r}$),
%($F_i(q)\equiv{\mathcal F}\{\rho_i(r)\}=\int\rho_i(r)\sin(qr)/(qr)d\pmb{r}$),
the relation can be written as
\begin{eqnarray}
%  F_{ch}(q) \approx F_p(q)G_E^p(q^2) + F_n(q)G_E^n(q^2) + {\rm ``spin-orbit''},
  F_{ch}(q) = F_p(q)G_E^p(q^2) + F_n(q)G_E^n(q^2) + F_{\rm SO}(q),
%  \\
%  \Leftrightarrow
%  F_p(q) &\approx& \frac{F_{ch}(q)-F_n(q)G_E^n(q^2)-F_{\rm SO}(q)}{G_E^p(q^2)},
  \label{eq:charge}
%  F_{\rm SO} &\approx&
%  -\frac{l(2j+1)(2G_M^n-G_E^n)}{4M^2}
%  -(2G_M^n-G_E^n)l(2j+1)/(2M)^2
%  {\mathcal F}\left\{\frac{\partial(r\rho_n^l(r))}{r^2\partial r}\right\}
%  \label{eq:so}
%  F_{ch}(q) \approx F_p(q)G_E^p(q^2) + F_n(q)G_E^n(q^2),
\end{eqnarray}
where $F_{ch}$, $F_{p(n)}$, and $G_{E}^{p(n)}$ are the nuclear charge,
the point-proton (neutron), and the Sachs single-nucleon
electric form factors, respectively.
%$G_E^{p(n)}$ was  defined as the so-called dipole form factor, which
The value of $r_{p(n)}^2=0.769$ fm$^2$ ($-0.116$ fm$^2$)
in Ref. \cite{Nakamura2010} is employed for 
the mean-square single-proton (neutron) charge radius of $G_E^{p(n)}$.
The nuclear charge distributions of $^{40,48}$Ca
were taken from Ref. \cite{DeVries1987c}.

The spin-orbit contribution $F_{\rm SO}$ is mainly attributed to
the nucleons in an open $l$ shell.
Although its contribution to the entire proton density is quite small
and neglected in many cases,
it has a non-negligible effect in the determination of $\Delta r_{np}$.
If $(2j+1)$ neutrons are filled in the $j=l+1/2$ sub-shell
and the proton shell is closed like $^{48}$Ca,
$F_{\rm SO}$ can be approximated as 
\begin{eqnarray}
  F_{\rm SO}(q) &\approx&
%  -\frac{l(2j+1)(2G_M^n(q^2)-G_E^n(q^2))}{4m^2}
  \frac{l(2j+1)(G_E^n(q^2)-2G_M^n(q^2))}{4m^2}
%  -(2G_M^n-G_E^n)l(2j+1)/(2m)^2
  {\mathcal F}\left\{\frac{\partial(r\rho_n^l(r))}{r^2\partial r}\right\},
  \label{eq:so}
%  F_{ch}(q) \approx F_p(q)G_E^p(q^2) + F_n(q)G_E^n(q^2),
\end{eqnarray}
where $G_M^n$ is the single-neutron magnetic form factor,
$\rho^l_n(r)$ is the density of each neutron in the $j=l+1/2$ subshell,
and $m$ is the nucleon mass.
(The relativistic description of $F_{\rm SO}$ is 
presented in Refs. \cite{Kurasawa2000, Horowitz2012a}.)
The spin-orbit contribution is sizable in $^{48}$Ca
due to the eight neutrons in the $f_{7/2}$ subshell.
The increase of $r_p$ in $^{48}$Ca due to the $F_{\rm SO}$ term
is evaluated to be $\sim$0.02 fm \cite{Bertozzi1972a, Horowitz2012a},
which is compatible with the typical uncertainty of $\Delta r_{np}$
\cite{Zenihiro2010}.
Hence, the spin-orbit contribution should be included to determine
%the proton density distribution
$\rho_p$ in $^{48}$Ca.
The $F_{\rm SO}$ contribution in $^{40}$Ca can be neglected
because the nucleons in the $LS$-closed shells do not contribute
to the $F_{\rm SO}$ term, as shown in Ref.~\cite{Bertozzi1972a}
for the non-relativistic limit.

\begin{figure}[t]
  \begin{center}
  \includegraphics[scale=0.29]{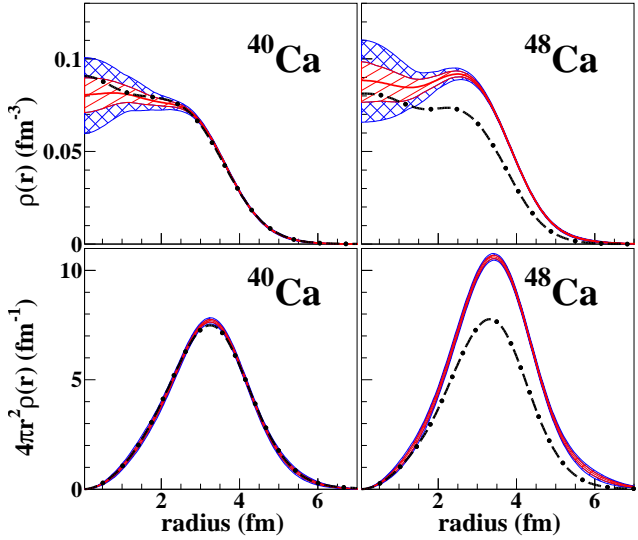}%
  \caption{
%    Extracted neutron density distributions with error-envelopes of
    Extracted $\rho_n$ with error-envelopes of
    $^{40,48}$Ca
    (red hatched and blue cross-hatched) and
%    the proton density distributions
    $\rho_p$
    derived from
    $\rho_{ch}$ (black dash-dotted).
    Upper panels are the density distribution ($\rho(r)$),
    and the lower panels 
    are the nucleon number distribution ($4\pi r^2\rho(r)$).
    \label{nden4048ca}}
  \end{center}
\end{figure}

The extraction of $\rho_n$ in $^{40,48}$Ca was carried out
by a $\chi^2$ fitting to the proton elastic scattering data
with $\rho_p$ determined by solving Eqs. (\ref{eq:charge}) and (\ref{eq:so})
iteratively.
%The neutron density distribution
$\rho_n$ is modeled
with a sum-of-Gaussian (SOG) function with 11 free parameters.
There is no {\it a priori} assumption on the form factor and
the extracted density distributions are independent of
specific nuclear structure models.
%The proton distribution
$\rho_p$ is derived by applying  $\rho_n$ determined from
the proton scattering data to the second and third terms
in Eq. (\ref{eq:charge}) and the  $\rho^l_n$ term in Eq. (\ref{eq:so}).
The $(2j+1)\rho^l_n$ is approximated with $\rho_n - \rho_p$.
The error of $\rho_p$ is not considered in this fitting procedure
since it is much smaller than that of $\rho_n$.
The $\chi^2$ fitting procedure started with initial values of
$\rho_n=(N/Z)\rho_p$ and $F_{\rm SO}=0$ and continued
until the self-consistent solution was obtained.

%The blue solid lines in Fig. \ref{data48ca} represent the predictions
%by the original MH model
%with Dirac-Hartree (DH) nucleon densities \cite{Horowitz1981a},
%while the red lines are the medium-modified
%RIA model with the best-fit density distributions discussed below.
%It is striking that both models reproduces the analyzing power data,
%but the density distributions should be changed
%to reproduce the cross section data.

While the blue solid lines in Fig. \ref{data48ca} represent the predictions
by the original MH model
with Dirac-Hartree (DH) nucleon densities \cite{Horowitz1981a},
the red solid lines are the best-fit
results with the reduced $\chi ^2$ minima 4.6 and 4.0 for $^{40,48}$Ca,
respectively.
%while the blue solid lines are due to the MH model calcualtions.
%The reduced $\chi ^2$ minima of $^{40,48}$Ca reached 4.6 and 4.0,
%respectively.
Figure \ref{nden4048ca} shows the extracted $\rho_n$ in $^{40,48}$Ca
(red solid) together with $\rho_p$ used in the search
after the iteration (black dash-dotted).
The upper panels in Fig. \ref{nden4048ca} show the density distributions,
whereas the lower panels
%are the nucleon number distributions ($4\pi r^2\rho(r)$).
are those multiplied by the phase space factor $4\pi r^2$.
%experimentally determined from the proton elastic scattering data
%The red solid lines in the middle of hatched areas are due to the
%best-fit result.
%with two types of error envelopes.
The red hatched areas show the standard error envelopes due to
the experimental statistical and systematic errors only.
The blue cross-hatched areas
are shown to visualize the maximum uncertainty of the present method
as well as the experimental errors.
The uncertainty is attributed to any effect that makes
the reduced $\chi^2$ larger than unity
and is evaluated by determining the density distributions
for the same data set but with artificially increased errors
so that the reduced $\chi^2$ becomes unity.
%Generally, the uncertainty due to the analysis model 
%is very difficult to evaluate and thus other works did not evaluate it.
The blue cross-hatched areas are comparable to
the red hatched areas. The results show that the present method is well 
established for the accurate determination of density distributions.
More details of the analysis method are reported
in Refs. \cite{Zenihiro2010, Sakaguchi2017}.
In $^{40}$Ca, $\rho_n$
%the extracted neutron density distribution
has almost the same shape as
$\rho_p$.
%the proton density distribution.
On the other hand,
%the neutron density distribution
$\rho_n$ in $^{48}$Ca 
is clearly enhanced over
$\rho_p$
%the proton density
and exhibits
the characteristic nose structure around 3 fm,
as shown in the top-right panel of Fig. \ref{nden4048ca}.
%, compared to that of $^{40}$Ca.
%This is mainly because the excess eight neutrons are occupied
%in the $1f_{7/2}$ orbit.
This structure is attributed to the radial distribution of the
$1f_{7/2}$ orbit in which the excess eight neutrons are filled.

\begin{table}[bt]
  \begin{center}
    \caption{
      Table of the rms radii and the skin thicknesses.
      $r_{ch}$ and $r_p$ used in this work,
      and the extracted $r_n$ and $\Delta r_{np}$
      in $^{40,48}$Ca are listed.
      $\delta^{\rm exp}$ and $\delta^{\rm exp+mdl}$
      are the two types of errors of $r_n$ and $\Delta r_{np}$
      due to the experimental errors only and the errors
      including model uncertainties, respectively.
      For $^{48}$Ca, some EDF and {\it ab initio} predictions
      are compared. All values are in fm.
      \label{radius}}
    \begin{ruledtabular}
      \begin{tabular}{l l c c c c c c}
%        & & $r_p$ & $r_n$ & $r_m$ & $\Delta r_{np}$
        & & $r_{ch}$ & $r_p$ & $r_n$ & $\Delta r_{np}$
%        & & $r_{ch}$ & $r_p$ & $r_n$
        & $\delta^{\rm exp}$ & $\delta^{\rm exp+mdl}$\\ \hline
        $^{40}$Ca
        & {\footnotesize This work} & 3.480 & 3.385 & 3.375 & $-0.010$
        & $^{+0.022}_{-0.023}$ & $^{+0.049}_{-0.048}$ \\
%        & {\footnotesize DD-MEb} & - & 3.37 & 3.32 & $-0.05$ & - & - \\
%        & {\footnotesize SAMi-J28} & - & 3.39 & 3.34 & $-0.05$ & - & - \\
%        & {\footnotesize $\Delta$NNLO} & - & 3.45 & 3.40 & $-0.05$ & - & - \\
        \hline
        $^{48}$Ca
        & {\footnotesize This work} & 3.460 & 3.387 & 3.555 & 0.168
        & $^{+0.025}_{-0.028}$ & $^{+0.052}_{-0.055}$ \\
%        & {\footnotesize DD-ME2} & - & 3.39 & 3.57 & 0.18 & - & - \\
        & {\footnotesize DD-MEb} & - & 3.39 & 3.57 & 0.18 & - & - \\
        & {\footnotesize SAMi-J28} & - & 3.44 & 3.60 & 0.16 & - & - \\
        & {\footnotesize NNLO$_{\rm sat}$} & - & 3.41 & 3.54 & 0.13 & - & - \\
        & {\footnotesize $\Delta$NNLO} & - & 3.47 & 3.62 & 0.15 & - & - \\
      \end{tabular}
    \end{ruledtabular}
  \end{center}
\end{table}

As listed in Table \ref{radius},
the neutron rms radii of $^{40,48}$Ca result in values of
3.375$^{+0.022}_{-0.023}$~fm
and 
3.555$^{+0.025}_{-0.028}$~fm, respectively.
The errors $\delta^{\rm exp}$ in Table \ref{radius}
correspond to the experimental error envelopes
(red hatched) in Fig. \ref{nden4048ca}.
If error envelopes which include the model ambiguities
(blue cross-hatched) are considered,
the $r_n$ values of $^{40,48}$Ca become
3.375$^{+0.049}_{-0.048}$~fm
and 
3.555$^{+0.052}_{-0.055}$~fm,
respectively.
%The first errors (``exp'') correspond to the experimental error envelopes
%(red hatched), while the second (``exp+mdl'') correspond to
%the error envelopes including the model uncertainties
%(blue cross-hatched) in Fig. \ref{nden4048ca}.
%The $\delta^{\rm exp}$ and $\delta^{\rm exp+mdl}$ errors show that
%the ambiguity due to the incompleteness of theory is much larger than
%the experimental error.
The $r_p$ values of $^{40,48}$Ca are
3.385 fm and 3.387 fm, respectively.
If $F_{\rm SO}$ in Eq. (\ref{eq:charge}) is not considered for $^{48}$Ca,
the $r_p$ value becomes 3.371 fm.
Table \ref{radius} lists the 
theoretical predictions of $r_p$ and $r_n$ of $^{48}$Ca
along with the experimental results:
the {\it ab initio} CC method with 
NNLO$_{\rm sat}$ and $\Delta$NNLO interactions
\cite{Ekstrom2015, Ekstrom2018}, as well as, 
a relativistic and a Skyrme EDF parameterization (DD-MEb, SAMi-J28)
of the Ref. \cite{Roca-Maza2013b}.
$\Delta$NNLO is the chiral EFT interaction
recently proposed by explicitly including $\Delta$-isobar
\cite{Ekstrom2018}.
NNLO$_{\rm sat}$ and DD-MEb give reasonable $r_p$ values
while $\Delta$NNLO and SAMi-J28 predict larger $r_p$ and $r_n$.
%Generally, it is difficult
%for EDF theories
%to reproduce the anomoulous behaviour of the charge (proton) radii
%in Ca isotopes,
%but the NNLO$_{\rm sat}$ prediction successfully reproduce the $r_{ch}$
%and $r_p$ values in both $^{40,48}$Ca \cite{Hagen2015}.
However, for the skin thickness of $^{48}$Ca, all the predictions
give consistent values with our result considering the error.
%Our result of $\Delta r_{np}$ is, however, consistent with those
%predictions within the error.

\begin{figure}[t]
  \begin{center}
  \includegraphics[scale=0.57]{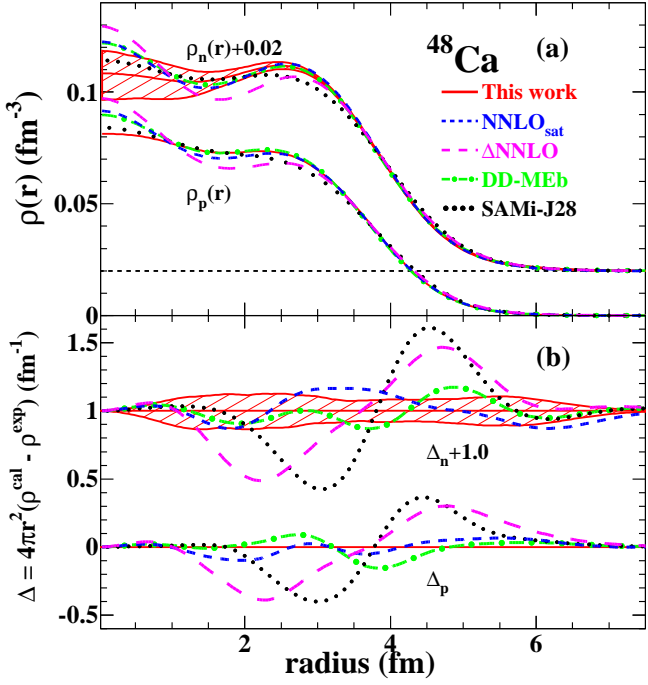}%
  \caption{
    (a) Extracted
%    neutron and proton density distributions
    $\rho_n$ and $\rho_p$
    of $^{48}$Ca
    compared with theoretical predictions.
    ``DD-MEb'' (green dash-dotted) and ``SAMi-J28'' (black dotted) are
    from a relativistic and a Skyrme EDF parameterizations, respectively,
    while ``NNLO$_{\rm sat}$'' (blue short-dash) and
    ``$\Delta$NNLO'' (magenta long-dash) are the predictions
    by the {\it ab initio} CC method.
%    (b) and (c) Density differences between
    (b) Density deviation ($4\pi r^2(\rho^{\rm cal}-\rho^{\rm exp})$) of 
    theories ($\rho^{\rm cal}$) presented in the upper panel (a)
    from our work ($\rho^{\rm exp}$).
%   and $\Delta\rho$ larger than 3.5~fm is expanded by five times in (c).
    \label{nden48ca3}}
  \end{center}
\end{figure}

Figure \ref{nden48ca3}
compares the experimentally-determined density distributions in $^{48}$Ca
(red solid and hatched lines) with
the predictions by NNLO$_{\rm sat}$ (blue short-dash),
$\Delta$NNLO (magenta long-dash),
DD-MEb (green dash-dotted),
and SAMi-J28 (black dotted). 
The lower panel (b) of Fig. \ref{nden48ca3} shows
the deviations of each prediction from our work
multiplied by $4\pi r^2$
($\Delta=4\pi r^2(\rho^{\rm cal}-\rho^{\rm exp})$).
%In Fig. \ref{nden48ca3}(c), at the radial region of larger than 3.5~fm,
%$\Delta\rho$ is multiplied by a factor of five.
For clarity,
$\rho_n$ and $\Delta_n$ are shifted by $+$0.02 fm$^{-3}$ and
$+$1.0 fm$^{-1}$, respectively.
The NNLO$_{\rm sat}$ and DD-MEb
predictions surprisingly agree with our result.
%, especially, in the surface region ($\geq 3$fm).
$\Delta$NNLO and SAMi-J28 predict large diffusenesses
of $\rho_p$ and $\rho_n$. Consequently, they
give larger $r_p$ and $r_n$, compared to our work.
%Different from NNLO$_{\rm sat}$ and DD-ME2,
%the diffusenesses of $\rho_p$ and $\rho_n$ predicted by SAMi-J28 are
%relatively large, and thus they give large $r_p$ and $r_n$.
It is thus demonstrated that comparison with the experimentally obtained
$\rho_p$ and $\rho_n$
%proton and neutron density distributions
enables the assessment of
the theoretical predictions which is not possible only with
the neutron skin thickness. This is an advantage in the proton
elastic scattering method over other methods that
determine only the radii or skin thicknesses.

\begin{figure}[t]
  \begin{center}
  \includegraphics[scale=0.24]{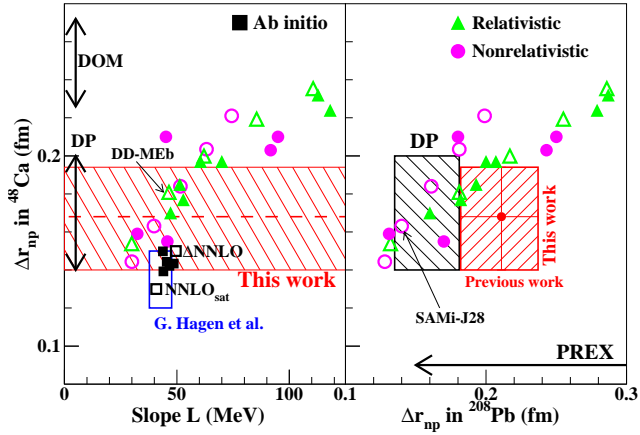}%
  \caption{
    (Left) $\Delta r_{np}$ in $^{48}$Ca versus the slope parameter $L$.
    (Right) $\Delta r_{np}$ in $^{48}$Ca versus $^{208}$Pb.
    Squares, triangles, and circles are predictions by
    {\it ab initio} CC method
    with chiral EFT interactions, relativistic and Skyrme EDFs,
    respectively.
    In the left panel, the arrows indicate the result of DP and
    DOM analyses, while the present result is shown by a red
    dash line with the hatched area. The blue rectangle shows
    the region evaluated in Ref. \cite{Hagen2015}.
    In the right panel, the red hatched region shows the overlap
    between this work for $^{48}$Ca and the previous work for
    $^{208}$Pb of Ref. \cite{Zenihiro2010}.
    The arrow is due to the PREX experiment,
    while the black hatched region is obtained by the DP analyses.
    \label{skinL48ca}}
  \end{center}
\end{figure}

%%%%%%%%%%%%%%%%%%%%%%%%%%%%%%%%%%%%%%%%%%%%%%%%%%%%%%%%%%%
%\paragraph*{Neutron skin thickness in $^{48}${\rm Ca}
%  and the slope parameter $L$}
The obtained values of $\Delta r_{np}$ in $^{40,48}$Ca
with the experimental errors are
$-0.010 ^{+0.022}_{-0.024}$ fm and $0.168 ^{+0.025}_{-0.028}$ fm,
respectively.
The $^{40}$Ca result is consistent with almost all the theoretical
calculations, which predict a small proton skin in $^{40}$Ca.
The small proton skin is a result of the repulsive Coulomb force
that pushes protons outwards.
%This means that our analysis method works well
% in the Ca isotope region.
The result of $\Delta r_{np}$ in $^{48}$Ca is consistent in
the range of 0.14--0.20 fm, which was recently obtained by 
interpreting the dipole polarizability (DP)
of $^{48}$Ca \cite{Birkhan2017}.
However, the value of $\Delta r_{np} = 0.249(23)$ fm
reported by the dispersive optical model (DOM) analysis
\cite{Mahzoon2017a} differs from our work and the DP result.
The left panel of Fig. \ref{skinL48ca} plots
the correlation between $\Delta r_{np}$ in $^{48}$Ca
and the slope parameter $L$ of the symmetry energy
predicted by the {\it ab initio}
and EDF models, while the right panel compares the
$\Delta r_{np}$ values of $^{48}$Ca and $^{208}$Pb
predicted by the relativistic and Skyrme EDF models.
Black squares, green triangles, and magenta circles
represent the predictions of the {\it ab initio} method,
the relativistic EDF models
(NL3 \cite{Lalazissis1997}, DD-ME2 \cite{Lalazissis2005}
DD-ME$\delta$ \cite{Roca-Maza2011a},
DD-PC1 \cite{Niksic2008c},
FSU \cite{Todd-Rutel2005a}, FSU2 \cite{Chen2014}
IUFSU \cite{Fattoyev2010})
and Skyrme EDF models
(SkM* \cite{Bartel1982a},
Sk255, Sk272 \cite{Agrawal2003},
SeaLL1 \cite{Bulgac2018},
UNEDF0 \cite{Kortelainen2010}),
respectively.
%\cite{Hagen2015, Ekstrom2018, Lalazissis2005, Agrawal2010,
%  Fattoyev2010, Kortelainen2010, Kortelainen2011, Bulgac2018}.
Open triangles and circles are sets of the DD-ME and SAMi-J families
in Ref. \cite{Roca-Maza2013b}, respectively.
Open squares are the {\it ab initio} results
by newly developed
NNLO$_{\rm sat}$ and $\Delta$NNLO interactions.
The blue rectangle represents the region
estimated from several chiral EFT interactions
by G. Hagen {\it et al.} in Ref. \cite{Hagen2015}.
The red hatched area shows the result of our work
%the extracted $\Delta r_{np}$ of $^{48}$Ca and
%its experimental error in this analysis,
while the arrows are the ranges by DP and DOM analyses.
%The EDF predictions in both left and righ show correlations
Compared to the recent EDF theories,
{\it ab initio} theories predict a slightly smaller neutron skin
thickness.
Our result for $^{48}$Ca implies
that $L$ is in the range of 20--70 MeV.
The right panel of Fig. \ref{skinL48ca} plots the correlation
between the $^{48}$Ca and $^{208}$Pb skin thicknesses.
A difference between the results from the proton elastic scattering
and from the dipole polarizability is noticeable.
The arrow denotes the PREX result of $^{208}$Pb \cite{Abrahamyan2012}.
%, and is still consistent with both results of DP and this work.

%%%%%%%%%%%%%%%%%%%%%%%%%%%%%%%%%%%%%%%%%%%%%%%%%%%%%%%%%%%%%%%%%%
In summary, we performed a direct determination of
the neutron density distributions and
the skin thicknesses of $^{40,48}$Ca from
proton elastic scattering at 295 MeV.
%, which was well established in heavier nulcei.
The obtained value of $\Delta r_{np} = 0.168^{+0.025}_{-0.028}$ fm
for $^{48}$Ca is consistent with the DP analysis, while
the DOM analysis provides a large skin thickness.
The recent {\it ab initio} and EDF predictions give
consistent values of $\Delta r_{np}$ with this work.
In particular, the calculations of the {\it ab initio} CC model
using the NNLO$_{\rm sat}$ interaction and the DD-MEb model 
provide density distributions that are consistent with our result.
%From the correlation between $\Delta r_{np}$ of $^{48}$Ca and $^{208}$Pb
%our combined results of proton elastic scattering method
%suggest different region by the DP analyses.

%%%ACKNOWLDGMENTS%%%
We would like to express our gratitude to the RCNP accelerator group
for providing the high-quality beam.
%We also thank
%G. Hagen, H. Sagawa, and S. Yoshida
%for providing valuable theoretical calculation results.
This work was in part supported by JSPS KAKENHI Grant Number 15H05451,
the U.S. Department of Energy, Office of Science,
Office of Nuclear Physics under Award Number
de-sc0018223 (SciDAC-4  NUCLEI), and the Field Work Proposals
ERKBP57and ERKBP72 at Oak Ridge National Laboratory(ORNL).
Computer time was provided by the Innovative and
Novel Computational Impact on Theory and
Experiment (INCITE) program.

%%%%%%%%%%%%%%%%%%%%%%%%%%%%%%%%%%%%%%%%%%%%%%%%%%%%%%%%%%%%%%%%%%%%%%%%%%%%%%

%\bibliography{bibliography/4048Ca}

\begin{thebibliography}{55}
\expandafter\ifx\csname natexlab\endcsname\relax\def\natexlab#1{#1}\fi
\expandafter\ifx\csname bibnamefont\endcsname\relax
  \def\bibnamefont#1{#1}\fi
\expandafter\ifx\csname bibfnamefont\endcsname\relax
  \def\bibfnamefont#1{#1}\fi
\expandafter\ifx\csname citenamefont\endcsname\relax
  \def\citenamefont#1{#1}\fi
\expandafter\ifx\csname url\endcsname\relax
  \def\url#1{\texttt{#1}}\fi
\expandafter\ifx\csname urlprefix\endcsname\relax\def\urlprefix{URL }\fi
\providecommand{\bibinfo}[2]{#2}
\providecommand{\eprint}[2][]{\url{#2}}

\bibitem[{\citenamefont{Oertel {\it et~al.}}(2017)\citenamefont{Oertel, Hempel,
  Kl{\"{a}}hn, and Typel}}]{Oertel2017}
\bibinfo{author}{\bibfnamefont{M.}~\bibnamefont{Oertel}}, \bibnamefont{{\it
  et~al.}}, \bibinfo{journal}{Rev. Mod. Phys.} \textbf{\bibinfo{volume}{89}},
  \bibinfo{pages}{015007} (\bibinfo{year}{2017}).

\bibitem[{\citenamefont{Radice {\it et~al.}}(2018)\citenamefont{Radice, Perego,
  Zappa, and Bernuzzi}}]{Radice2017}
\bibinfo{author}{\bibfnamefont{D.}~\bibnamefont{Radice}}, \bibnamefont{{\it
  et~al.}}, \bibinfo{journal}{Astrophys. J.} \textbf{\bibinfo{volume}{852}},
  \bibinfo{pages}{L29} (\bibinfo{year}{2018}).

\bibitem[{\citenamefont{Roca-Maza and Paar}(2018)}]{Roca-Maza2018}
\bibinfo{author}{\bibfnamefont{X.}~\bibnamefont{Roca-Maza}} \bibnamefont{and}
  \bibinfo{author}{\bibfnamefont{N.}~\bibnamefont{Paar}},
  \bibinfo{journal}{Prog. Part. Nucl. Phys.} \textbf{\bibinfo{volume}{101}},
  \bibinfo{pages}{96} (\bibinfo{year}{2018}).

\bibitem[{\citenamefont{Fattoyev {\it et~al.}}(2018)\citenamefont{Fattoyev,
  Piekarewicz, and Horowitz}}]{Fattoyev2018}
\bibinfo{author}{\bibfnamefont{F.~J.} \bibnamefont{Fattoyev}},
  \bibinfo{author}{\bibfnamefont{J.}~\bibnamefont{Piekarewicz}},
  \bibnamefont{and} \bibinfo{author}{\bibfnamefont{C.~J.}
  \bibnamefont{Horowitz}}, \bibinfo{journal}{Phys. Rev. Lett.}
  \textbf{\bibinfo{volume}{120}}, \bibinfo{pages}{172702}
  (\bibinfo{year}{2018}).

\bibitem[{\citenamefont{Annala {\it et~al.}}(2018)\citenamefont{Annala, Gorda,
  Kurkela, and Vuorinen}}]{Annala2018}
\bibinfo{author}{\bibfnamefont{E.}~\bibnamefont{Annala}}, \bibnamefont{{\it
  et~al.}}, \bibinfo{journal}{Phys. Rev. Lett.} \textbf{\bibinfo{volume}{120}},
  \bibinfo{pages}{172703} (\bibinfo{year}{2018}).

\bibitem[{\citenamefont{Abbott {\it et~al.}}(2017)\citenamefont{Abbott, Abbott,
  Abbott, Acernese, Ackley, Adams, Adams, Addesso, Adhikari, Adya {\it
  et~al.}}}]{Abbott2017}
\bibinfo{author}{\bibfnamefont{B.~P.} \bibnamefont{Abbott}}, \bibnamefont{{\it
  et~al.}}, \bibinfo{journal}{Phys. Rev. Lett.} \textbf{\bibinfo{volume}{119}},
  \bibinfo{pages}{30} (\bibinfo{year}{2017}).

\bibitem[{\citenamefont{Typel and Brown}(2001)}]{Typel2001}
\bibinfo{author}{\bibfnamefont{S.}~\bibnamefont{Typel}} \bibnamefont{and}
  \bibinfo{author}{\bibfnamefont{B.~A.} \bibnamefont{Brown}},
  \bibinfo{journal}{Phys. Rev. C} \textbf{\bibinfo{volume}{64}},
  \bibinfo{pages}{027302} (\bibinfo{year}{2001}).

\bibitem[{\citenamefont{Furnstahl}(2002)}]{Furnstahl2002}
\bibinfo{author}{\bibfnamefont{R.~J.} \bibnamefont{Furnstahl}},
  \bibinfo{journal}{Nucl. Phys. A} \textbf{\bibinfo{volume}{706}},
  \bibinfo{pages}{85} (\bibinfo{year}{2002}).

\bibitem[{\citenamefont{Chen {\it et~al.}}(2005)\citenamefont{Chen, Ko, and
  Li}}]{Chen2005a}
\bibinfo{author}{\bibfnamefont{L.-W.} \bibnamefont{Chen}},
  \bibinfo{author}{\bibfnamefont{C.~M.} \bibnamefont{Ko}}, \bibnamefont{and}
  \bibinfo{author}{\bibfnamefont{B.-A.} \bibnamefont{Li}},
  \bibinfo{journal}{Phys. Rev. C} \textbf{\bibinfo{volume}{72}},
  \bibinfo{pages}{064309} (\bibinfo{year}{2005}).

\bibitem[{\citenamefont{Starodubsky and Hintz}(1994)}]{Starodubsky1994b}
\bibinfo{author}{\bibfnamefont{V.~E.} \bibnamefont{Starodubsky}}
  \bibnamefont{and} \bibinfo{author}{\bibfnamefont{N.~M.} \bibnamefont{Hintz}},
  \bibinfo{journal}{Phys. Rev. C} \textbf{\bibinfo{volume}{49}},
  \bibinfo{pages}{2118} (\bibinfo{year}{1994}).

\bibitem[{\citenamefont{Zenihiro {\it et~al.}}(2010)\citenamefont{Zenihiro,
  Sakaguchi, Murakami, Yosoi, Yasuda, Terashima, Iwao, Takeda, Itoh, Yoshida
  {\it et~al.}}}]{Zenihiro2010}
\bibinfo{author}{\bibfnamefont{J.}~\bibnamefont{Zenihiro}}, \bibnamefont{{\it
  et~al.}}, \bibinfo{journal}{Phys. Rev. C} \textbf{\bibinfo{volume}{82}},
  \bibinfo{pages}{44611} (\bibinfo{year}{2010}).

\bibitem[{\citenamefont{Tarbert {\it et~al.}}(2014)\citenamefont{Tarbert,
  Watts, Glazier, Aguar, Ahrens, Annand, Arends, Beck, Bekrenev, Boillat {\it
  et~al.}}}]{Tarbert2014a}
\bibinfo{author}{\bibfnamefont{C.~M.} \bibnamefont{Tarbert}}, \bibnamefont{{\it
  et~al.}}, \bibinfo{journal}{Phys. Rev. Lett.} \textbf{\bibinfo{volume}{112}},
  \bibinfo{pages}{242502} (\bibinfo{year}{2014}).

\bibitem[{\citenamefont{K{\l}os {\it et~al.}}(2007)\citenamefont{K{\l}os,
  Trzci{\'{n}}ska, Jastrz{\c{e}}bski, Czosnyka, Kisieli{\'{n}}ski,
  Lubi{\'{n}}ski, Napiorkowski, Pie{\'{n}}kowski, Hartmann, Ketzer {\it
  et~al.}}}]{Kos2007}
\bibinfo{author}{\bibfnamefont{B.}~\bibnamefont{K{\l}os}}, \bibnamefont{{\it
  et~al.}}, \bibinfo{journal}{Phys. Rev. C} \textbf{\bibinfo{volume}{76}},
  \bibinfo{pages}{014311} (\bibinfo{year}{2007}).

\bibitem[{\citenamefont{Tamii {\it et~al.}}(2011)\citenamefont{Tamii,
  Poltoratska, {Von Neumann-Cosel}, Fujita, Adachi, Bertulani, Carter, Dozono,
  Fujita, Fujita {\it et~al.}}}]{Tamii2011a}
\bibinfo{author}{\bibfnamefont{A.}~\bibnamefont{Tamii}}, \bibnamefont{{\it
  et~al.}}, \bibinfo{journal}{Phys. Rev. Lett.} \textbf{\bibinfo{volume}{107}},
  \bibinfo{pages}{1} (\bibinfo{year}{2011}).

\bibitem[{\citenamefont{Abrahamyan {\it et~al.}}(2012)\citenamefont{Abrahamyan,
  Ahmed, Albataineh, Aniol, Armstrong, Armstrong, Averett, Babineau, Barbieri,
  Bellini {\it et~al.}}}]{Abrahamyan2012}
\bibinfo{author}{\bibfnamefont{S.}~\bibnamefont{Abrahamyan}}, \bibnamefont{{\it
  et~al.}}, \bibinfo{journal}{Phys. Rev. Lett.} \textbf{\bibinfo{volume}{108}},
  \bibinfo{pages}{112502} (\bibinfo{year}{2012}).

\bibitem[{\citenamefont{Roca-Maza {\it
  et~al.}}(2011{\natexlab{a}})\citenamefont{Roca-Maza, Centelles, Vi{\~{n}}as,
  and Warda}}]{Roca-Maza2011}
\bibinfo{author}{\bibfnamefont{X.}~\bibnamefont{Roca-Maza}}, \bibnamefont{{\it
  et~al.}}, \bibinfo{journal}{Phys. Rev. Lett.} \textbf{\bibinfo{volume}{106}},
  \bibinfo{pages}{252501} (\bibinfo{year}{2011}{\natexlab{a}}).

\bibitem[{\citenamefont{Tsang {\it et~al.}}(2012)\citenamefont{Tsang, Stone,
  Camera, Danielewicz, Gandolfi, Hebeler, Horowitz, Lee, Lynch, Kohley {\it
  et~al.}}}]{Tsang2012}
\bibinfo{author}{\bibfnamefont{M.~B.} \bibnamefont{Tsang}}, \bibnamefont{{\it
  et~al.}}, \bibinfo{journal}{Phys. Rev. C} \textbf{\bibinfo{volume}{86}},
  \bibinfo{pages}{015803} (\bibinfo{year}{2012}).

\bibitem[{\citenamefont{Furnstahl and Hebeler}(2013)}]{Furnstahl2013}
\bibinfo{author}{\bibfnamefont{R.~J.} \bibnamefont{Furnstahl}}
  \bibnamefont{and} \bibinfo{author}{\bibfnamefont{K.}~\bibnamefont{Hebeler}},
  \bibinfo{journal}{Reports Prog. Phys.} \textbf{\bibinfo{volume}{76}},
  \bibinfo{pages}{126301} (\bibinfo{year}{2013}).

\bibitem[{\citenamefont{Som{\`{a}} {\it et~al.}}(2014)\citenamefont{Som{\`{a}},
  Cipollone, Barbieri, Navr{\'{a}}til, and Duguet}}]{Soma2014}
\bibinfo{author}{\bibfnamefont{V.}~\bibnamefont{Som{\`{a}}}}, \bibnamefont{{\it
  et~al.}}, \bibinfo{journal}{Phys. Rev. C} \textbf{\bibinfo{volume}{89}},
  \bibinfo{pages}{061301} (\bibinfo{year}{2014}).

\bibitem[{\citenamefont{Ekstr{\"{o}}m {\it
  et~al.}}(2015)\citenamefont{Ekstr{\"{o}}m, Jansen, Wendt, Hagen, Papenbrock,
  Carlsson, Forss{\'{e}}n, Hjorth-Jensen, Navr{\'{a}}til, and
  Nazarewicz}}]{Ekstrom2015}
\bibinfo{author}{\bibfnamefont{A.}~\bibnamefont{Ekstr{\"{o}}m}},
  \bibnamefont{{\it et~al.}}, \bibinfo{journal}{Phys. Rev. C}
  \textbf{\bibinfo{volume}{91}}, \bibinfo{pages}{051301}
  (\bibinfo{year}{2015}).

\bibitem[{\citenamefont{Hergert {\it et~al.}}(2016)\citenamefont{Hergert,
  Bogner, Morris, Schwenk, and Tsukiyama}}]{Hergert2016}
\bibinfo{author}{\bibfnamefont{H.}~\bibnamefont{Hergert}}, \bibnamefont{{\it
  et~al.}}, \bibinfo{journal}{Phys. Rep.} \textbf{\bibinfo{volume}{621}},
  \bibinfo{pages}{165} (\bibinfo{year}{2016}).

\bibitem[{\citenamefont{Ekstr{\"{o}}m {\it
  et~al.}}(2018)\citenamefont{Ekstr{\"{o}}m, Hagen, Morris, Papenbrock, and
  Schwartz}}]{Ekstrom2018}
\bibinfo{author}{\bibfnamefont{A.}~\bibnamefont{Ekstr{\"{o}}m}},
  \bibnamefont{{\it et~al.}}, \bibinfo{journal}{Phys. Rev. C}
  \textbf{\bibinfo{volume}{97}}, \bibinfo{pages}{024332}
  (\bibinfo{year}{2018}).

\bibitem[{\citenamefont{Akmal {\it et~al.}}(1998)\citenamefont{Akmal,
  Pandharipande, and Ravenhall}}]{Akmal1998}
\bibinfo{author}{\bibfnamefont{A.}~\bibnamefont{Akmal}},
  \bibinfo{author}{\bibfnamefont{V.~R.} \bibnamefont{Pandharipande}},
  \bibnamefont{and} \bibinfo{author}{\bibfnamefont{D.~G.}
  \bibnamefont{Ravenhall}}, \bibinfo{journal}{Phys. Rev. C}
  \textbf{\bibinfo{volume}{58}}, \bibinfo{pages}{1804} (\bibinfo{year}{1998}).

\bibitem[{\citenamefont{Hagen {\it et~al.}}(2016)\citenamefont{Hagen,
  Hjorth-Jensen, Jansen, and Papenbrock}}]{Hagen2016a}
\bibinfo{author}{\bibfnamefont{G.}~\bibnamefont{Hagen}}, \bibnamefont{{\it
  et~al.}}, \bibinfo{journal}{Phys. Scr.} \textbf{\bibinfo{volume}{91}},
  \bibinfo{pages}{063006} (\bibinfo{year}{2016}).

\bibitem[{\citenamefont{Hagen {\it et~al.}}(2015)\citenamefont{Hagen,
  Ekstr{\"{o}}m, Forss{\'{e}}n, Jansen, Nazarewicz, Papenbrock, Wendt, Bacca,
  Barnea, Carlsson {\it et~al.}}}]{Hagen2015}
\bibinfo{author}{\bibfnamefont{G.}~\bibnamefont{Hagen}}, \bibnamefont{{\it
  et~al.}}, \bibinfo{journal}{Nat. Phys.} \textbf{\bibinfo{volume}{12}},
  \bibinfo{pages}{186} (\bibinfo{year}{2015}).

\bibitem[{\citenamefont{Birkhan {\it et~al.}}(2017)\citenamefont{Birkhan,
  Miorelli, Bacca, Bassauer, Bertulani, Hagen, Matsubara, {Von Neumann-Cosel},
  Papenbrock, Pietralla {\it et~al.}}}]{Birkhan2017}
\bibinfo{author}{\bibfnamefont{J.}~\bibnamefont{Birkhan}}, \bibnamefont{{\it
  et~al.}}, \bibinfo{journal}{Phys. Rev. Lett.} \textbf{\bibinfo{volume}{118}},
  \bibinfo{pages}{252501} (\bibinfo{year}{2017}).

\bibitem[{\citenamefont{Horowitz {\it et~al.}}(2014)\citenamefont{Horowitz,
  Kumar, and Michaels}}]{Horowitz2014}
\bibinfo{author}{\bibfnamefont{C.~J.} \bibnamefont{Horowitz}},
  \bibinfo{author}{\bibfnamefont{K.~S.} \bibnamefont{Kumar}}, \bibnamefont{and}
  \bibinfo{author}{\bibfnamefont{R.}~\bibnamefont{Michaels}},
  \bibinfo{journal}{Eur. Phys. J. A} \textbf{\bibinfo{volume}{50}},
  \bibinfo{pages}{48} (\bibinfo{year}{2014}).

\bibitem[{\citenamefont{Terashima {\it et~al.}}(2008)\citenamefont{Terashima,
  Sakaguchi, Takeda, Ishikawa, Itoh, Kawabata, Murakami, Uchida, Yasuda, Yosoi
  {\it et~al.}}}]{Terashima2008}
\bibinfo{author}{\bibfnamefont{S.}~\bibnamefont{Terashima}}, \bibnamefont{{\it
  et~al.}}, \bibinfo{journal}{Phys. Rev. C} \textbf{\bibinfo{volume}{77}},
  \bibinfo{pages}{024317} (\bibinfo{year}{2008}).

\bibitem[{\citenamefont{Fujiwara {\it et~al.}}(1999)\citenamefont{Fujiwara,
  Akimune, Daito, Fujimura, Fujita, Hatanaka, Ikegami, Katayama, Nagayama,
  Matsuoka {\it et~al.}}}]{Fujiwara1999}
\bibinfo{author}{\bibfnamefont{M.}~\bibnamefont{Fujiwara}}, \bibnamefont{{\it
  et~al.}}, \bibinfo{journal}{Nucl. Instruments Methods Phys. Res. Sect. A
  Accel. Spectrometers, Detect. Assoc. Equip.} \textbf{\bibinfo{volume}{422}},
  \bibinfo{pages}{484} (\bibinfo{year}{1999}).

\bibitem[{\citenamefont{Sakaguchi and Zenihiro}(2017)}]{Sakaguchi2017}
\bibinfo{author}{\bibfnamefont{H.}~\bibnamefont{Sakaguchi}} \bibnamefont{and}
  \bibinfo{author}{\bibfnamefont{J.}~\bibnamefont{Zenihiro}},
  \bibinfo{journal}{Prog. Part. Nucl. Phys.} \textbf{\bibinfo{volume}{97}},
  \bibinfo{pages}{1} (\bibinfo{year}{2017}).

\bibitem[{\citenamefont{Murdock and
  Horowitz}(1987)}]{MurdockDavidPandHorowitz1987}
\bibinfo{author}{\bibfnamefont{D.~P.} \bibnamefont{Murdock}} \bibnamefont{and}
  \bibinfo{author}{\bibfnamefont{C.~J.} \bibnamefont{Horowitz}},
  \bibinfo{journal}{Phys. Rev. C} \textbf{\bibinfo{volume}{35}},
  \bibinfo{pages}{1442} (\bibinfo{year}{1987}).

\bibitem[{\citenamefont{Horowitz}(1985)}]{Horowitz1985}
\bibinfo{author}{\bibfnamefont{C.~J.} \bibnamefont{Horowitz}},
  \bibinfo{journal}{Phys. Rev. C} \textbf{\bibinfo{volume}{31}},
  \bibinfo{pages}{1340} (\bibinfo{year}{1985}).

\bibitem[{\citenamefont{Sakaguchi {\it et~al.}}(1998)\citenamefont{Sakaguchi,
  Takeda, Toyama, Itoh, Yamagoshi, Tamii, Yosoi, Akimune, Daito, Inomata {\it
  et~al.}}}]{Sakaguchi1998}
\bibinfo{author}{\bibfnamefont{H.}~\bibnamefont{Sakaguchi}}, \bibnamefont{{\it
  et~al.}}, \bibinfo{journal}{Phys. Rev. C} \textbf{\bibinfo{volume}{57}},
  \bibinfo{pages}{1749} (\bibinfo{year}{1998}).

\bibitem[{\citenamefont{{De Vries} {\it et~al.}}(1987)\citenamefont{{De Vries},
  {De Jager}, and {De Vries}}}]{DeVries1987c}
\bibinfo{author}{\bibfnamefont{H.}~\bibnamefont{{De Vries}}},
  \bibinfo{author}{\bibfnamefont{C.}~\bibnamefont{{De Jager}}},
  \bibnamefont{and} \bibinfo{author}{\bibfnamefont{C.}~\bibnamefont{{De
  Vries}}}, \bibinfo{journal}{At. Data Nucl. Data Tables}
  \textbf{\bibinfo{volume}{36}}, \bibinfo{pages}{495} (\bibinfo{year}{1987}).

%\bibitem[{\citenamefont{Hofmann {\it et~al.}}(2001)\citenamefont{Hofmann, Keil,
%  and Lenske}}]{Hofmann2001}
%\bibinfo{author}{\bibfnamefont{F.}~\bibnamefont{Hofmann}},
%  \bibinfo{author}{\bibfnamefont{C.~M.} \bibnamefont{Keil}}, \bibnamefont{and}
%  \bibinfo{author}{\bibfnamefont{H.}~\bibnamefont{Lenske}},
%  \bibinfo{journal}{Phys. Rev. C} \textbf{\bibinfo{volume}{64}},
%  \bibinfo{pages}{034314} (\bibinfo{year}{2001}).

%\bibitem[{\citenamefont{Vretenar {\it et~al.}}(2005)\citenamefont{Vretenar,
%  Afanasjev, Lalazissis, and Ring}}]{VRETENAR2005a}
%\bibinfo{author}{\bibfnamefont{D.}~\bibnamefont{Vretenar}}, \bibnamefont{{\it
%  et~al.}}, \bibinfo{journal}{Phys. Rep.} \textbf{\bibinfo{volume}{409}},
%  \bibinfo{pages}{101} (\bibinfo{year}{2005}).

%\bibitem[{\citenamefont{Sarriguren {\it et~al.}}(2007)\citenamefont{Sarriguren,
%  Gaidarov, de~Guerra, and Antonov}}]{Sarriguren2007}
%\bibinfo{author}{\bibfnamefont{P.}~\bibnamefont{Sarriguren}}, \bibnamefont{{\it
%  et~al.}}, \bibinfo{journal}{Phys. Rev. C} \textbf{\bibinfo{volume}{76}},
%  \bibinfo{pages}{044322} (\bibinfo{year}{2007}).

\bibitem[{\citenamefont{Horowitz and Serot}(1981)}]{Horowitz1981a}
\bibinfo{author}{\bibfnamefont{C.~J.} \bibnamefont{Horowitz}} \bibnamefont{and}
  \bibinfo{author}{\bibfnamefont{B.~D.} \bibnamefont{Serot}},
  \bibinfo{journal}{Nucl. Physics, Sect. A} \textbf{\bibinfo{volume}{368}},
  \bibinfo{pages}{503} (\bibinfo{year}{1981}).

\bibitem[{\citenamefont{Nakamura {\it et~al.}}(2010)\citenamefont{Nakamura,
  Hagiwara, Hikasa, Murayama, Tanabashi, Watari, Amsler, Antonelli, Asner, Baer
  {\it et~al.}}}]{Nakamura2010}
\bibinfo{author}{\bibfnamefont{K.}~\bibnamefont{Nakamura}}, \bibnamefont{{\it
  et~al.}}, \bibinfo{journal}{J. Phys. G Nucl. Part. Phys.}
  \textbf{\bibinfo{volume}{37}} (\bibinfo{year}{2010}).

\bibitem[{\citenamefont{Kurasawa and Suzuki}(2000)}]{Kurasawa2000}
\bibinfo{author}{\bibfnamefont{H.}~\bibnamefont{Kurasawa}} \bibnamefont{and}
  \bibinfo{author}{\bibfnamefont{T.}~\bibnamefont{Suzuki}},
  \bibinfo{journal}{Phys. Rev. C} \textbf{\bibinfo{volume}{62}},
  \bibinfo{pages}{054303} (\bibinfo{year}{2000}).

\bibitem[{\citenamefont{Horowitz and Piekarewicz}(2012)}]{Horowitz2012a}
\bibinfo{author}{\bibfnamefont{C.~J.} \bibnamefont{Horowitz}} \bibnamefont{and}
  \bibinfo{author}{\bibfnamefont{J.}~\bibnamefont{Piekarewicz}},
  \bibinfo{journal}{Phys. Rev. C} \textbf{\bibinfo{volume}{86}},
  \bibinfo{pages}{045503} (\bibinfo{year}{2012}).

\bibitem[{\citenamefont{Bertozzi {\it et~al.}}(1972)\citenamefont{Bertozzi,
  Friar, Heisenberg, and Negele}}]{Bertozzi1972a}
\bibinfo{author}{\bibfnamefont{W.}~\bibnamefont{Bertozzi}}, \bibnamefont{{\it
  et~al.}}, \bibinfo{journal}{Phys. Lett. B} \textbf{\bibinfo{volume}{41}},
  \bibinfo{pages}{408} (\bibinfo{year}{1972}).

\bibitem[{\citenamefont{Roca-Maza {\it et~al.}}(2013)\citenamefont{Roca-Maza,
  Brenna, Agrawal, Bortignon, Col{\`{o}}, Cao, Paar, and
  Vretenar}}]{Roca-Maza2013b}
\bibinfo{author}{\bibfnamefont{X.}~\bibnamefont{Roca-Maza}}, \bibnamefont{{\it
  et~al.}}, \bibinfo{journal}{Phys. Rev. C} \textbf{\bibinfo{volume}{87}},
  \bibinfo{pages}{034301} (\bibinfo{year}{2013}).

\bibitem[{\citenamefont{Mahzoon {\it et~al.}}(2017)\citenamefont{Mahzoon,
  Atkinson, Charity, and Dickhoff}}]{Mahzoon2017a}
\bibinfo{author}{\bibfnamefont{M.~H.} \bibnamefont{Mahzoon}}, \bibnamefont{{\it
  et~al.}}, \bibinfo{journal}{Phys. Rev. Lett.} \textbf{\bibinfo{volume}{119}},
  \bibinfo{pages}{222503} (\bibinfo{year}{2017}).

\bibitem[{\citenamefont{Lalazissis {\it et~al.}}(1997)\citenamefont{Lalazissis,
  K{\"{o}}nig, and Ring}}]{Lalazissis1997}
\bibinfo{author}{\bibfnamefont{G.~a.} \bibnamefont{Lalazissis}},
  \bibinfo{author}{\bibfnamefont{J.}~\bibnamefont{K{\"{o}}nig}},
  \bibnamefont{and} \bibinfo{author}{\bibfnamefont{P.}~\bibnamefont{Ring}},
  \bibinfo{journal}{Phys. Rev. C} \textbf{\bibinfo{volume}{55}},
  \bibinfo{pages}{540} (\bibinfo{year}{1997}).

\bibitem[{\citenamefont{Lalazissis {\it et~al.}}(2005)\citenamefont{Lalazissis,
  Nik{\v{s}}i{\'{c}}, Vretenar, and Ring}}]{Lalazissis2005}
\bibinfo{author}{\bibfnamefont{G.~A.} \bibnamefont{Lalazissis}},
  \bibnamefont{{\it et~al.}}, \bibinfo{journal}{Phys. Rev. C}
  \textbf{\bibinfo{volume}{71}}, \bibinfo{pages}{024312}
  (\bibinfo{year}{2005}).

\bibitem[{\citenamefont{Roca-Maza {\it
  et~al.}}(2011{\natexlab{b}})\citenamefont{Roca-Maza, Vi{\~{n}}as, Centelles,
  Ring, and Schuck}}]{Roca-Maza2011a}
\bibinfo{author}{\bibfnamefont{X.}~\bibnamefont{Roca-Maza}}, \bibnamefont{{\it
  et~al.}}, \bibinfo{journal}{Phys. Rev. C} \textbf{\bibinfo{volume}{84}},
  \bibinfo{pages}{054309} (\bibinfo{year}{2011}{\natexlab{b}}).

\bibitem[{\citenamefont{Nik{\v{s}}i{\'{c}} {\it
  et~al.}}(2008)\citenamefont{Nik{\v{s}}i{\'{c}}, Vretenar, and
  Ring}}]{Niksic2008c}
\bibinfo{author}{\bibfnamefont{T.}~\bibnamefont{Nik{\v{s}}i{\'{c}}}},
  \bibinfo{author}{\bibfnamefont{D.}~\bibnamefont{Vretenar}}, \bibnamefont{and}
  \bibinfo{author}{\bibfnamefont{P.}~\bibnamefont{Ring}},
  \bibinfo{journal}{Phys. Rev. C} \textbf{\bibinfo{volume}{78}},
  \bibinfo{pages}{034318} (\bibinfo{year}{2008}).

\bibitem[{\citenamefont{Todd-Rutel and Piekarewicz}(2005)}]{Todd-Rutel2005a}
\bibinfo{author}{\bibfnamefont{B.~G.} \bibnamefont{Todd-Rutel}}
  \bibnamefont{and}
  \bibinfo{author}{\bibfnamefont{J.}~\bibnamefont{Piekarewicz}},
  \bibinfo{journal}{Phys. Rev. Lett.} \textbf{\bibinfo{volume}{95}},
  \bibinfo{pages}{1} (\bibinfo{year}{2005}).

\bibitem[{\citenamefont{Chen and Piekarewicz}(2014)}]{Chen2014}
\bibinfo{author}{\bibfnamefont{W.-C.} \bibnamefont{Chen}} \bibnamefont{and}
  \bibinfo{author}{\bibfnamefont{J.}~\bibnamefont{Piekarewicz}},
  \bibinfo{journal}{Phys. Rev. C} \textbf{\bibinfo{volume}{90}},
  \bibinfo{pages}{044305} (\bibinfo{year}{2014}).

\bibitem[{\citenamefont{Fattoyev {\it et~al.}}(2010)\citenamefont{Fattoyev,
  Horowitz, Piekarewicz, and Shen}}]{Fattoyev2010}
\bibinfo{author}{\bibfnamefont{F.~J.} \bibnamefont{Fattoyev}},
  \bibnamefont{{\it et~al.}}, \bibinfo{journal}{Phys. Rev. C}
  \textbf{\bibinfo{volume}{82}}, \bibinfo{pages}{055803}
  (\bibinfo{year}{2010}).

\bibitem[{\citenamefont{Bartel {\it et~al.}}(1982)\citenamefont{Bartel,
  Quentin, Brack, Guet, and H{\aa}kansson}}]{Bartel1982a}
\bibinfo{author}{\bibfnamefont{J.}~\bibnamefont{Bartel}}, \bibnamefont{{\it
  et~al.}}, \bibinfo{journal}{Nucl. Phys. A} \textbf{\bibinfo{volume}{386}},
  \bibinfo{pages}{79} (\bibinfo{year}{1982}).

\bibitem[{\citenamefont{Agrawal {\it et~al.}}(2003)\citenamefont{Agrawal,
  Shlomo, and {Kim Au}}}]{Agrawal2003}
\bibinfo{author}{\bibfnamefont{B.~K.} \bibnamefont{Agrawal}},
  \bibinfo{author}{\bibfnamefont{S.}~\bibnamefont{Shlomo}}, \bibnamefont{and}
  \bibinfo{author}{\bibfnamefont{V.}~\bibnamefont{{Kim Au}}},
  \bibinfo{journal}{Phys. Rev. C} \textbf{\bibinfo{volume}{68}},
  \bibinfo{pages}{031304} (\bibinfo{year}{2003}).

\bibitem[{\citenamefont{Bulgac {\it et~al.}}(2018)\citenamefont{Bulgac, Forbes,
  Jin, Perez, and Schunck}}]{Bulgac2018}
\bibinfo{author}{\bibfnamefont{A.}~\bibnamefont{Bulgac}}, \bibnamefont{{\it
  et~al.}}, \bibinfo{journal}{Phys. Rev. C} \textbf{\bibinfo{volume}{97}},
  \bibinfo{pages}{044313} (\bibinfo{year}{2018}).

\bibitem[{\citenamefont{Kortelainen {\it
  et~al.}}(2010)\citenamefont{Kortelainen, Lesinski, Mor{\'{e}}, Nazarewicz,
  Sarich, Schunck, Stoitsov, and Wild}}]{Kortelainen2010}
\bibinfo{author}{\bibfnamefont{M.}~\bibnamefont{Kortelainen}},
  \bibnamefont{{\it et~al.}}, \bibinfo{journal}{Phys. Rev. C}
  \textbf{\bibinfo{volume}{82}}, \bibinfo{pages}{024313}
  (\bibinfo{year}{2010}).

\end{thebibliography}

%%%%%%%%%%%%

\end{document}